\documentclass[aps,showpacs,showkeys,preprint,prb]{revtex4-1}
\usepackage{graphicx}

\usepackage{amssymb}
\usepackage{siunitx}
\usepackage[utf8]{inputenc}
\usepackage{color}

\begin{document}

\title{From soft to hard magnetic Fe-Co-B by spontaneous strain: A combined first principle and thin film study}

\author{L. Reichel}
\email[Corresponding author: ]{l.reichel@ifw-dresden.de}
\author{L. Schultz}
\affiliation{IFW Dresden, P.O. Box 270116, 01171 Dresden, Germany}
\affiliation{TU Dresden, Faculty of Mechanical Engineering, Institute of Materials Science, 01062 Dresden, Germany}
\author{D. Pohl}
\author{S. Oswald}
\author{S. F\"ahler}
\affiliation{IFW Dresden, P.O. Box 270116, 01171 Dresden, Germany}
\author{M. Werwi\'{n}ski}
\affiliation{Division of Materials Theory, Department of Physics and Astronomy, Uppsala University, Box 516, SE-751 20, Uppsala, Sweden}
\affiliation{Institute of Molecular Physics, Polish Academy of Sciences, 60-179 Poznań, Poland}
\author{A. Edstr\"om}
\author{E. K. Delczeg-Czirjak}
\author{J. Rusz}
\affiliation{Division of Materials Theory, Department of Physics and Astronomy, Uppsala University, Box 516, SE-751 20, Uppsala, Sweden}

\begin{abstract}

In order to convert the well-known Fe-Co-B alloy from a soft to a hard magnet, we propose tetragonal strain by interstitial boron. Density functional theory reveals that when B atoms occupy octahedral interstitial sites, the bcc Fe-Co lattice is strained spontaneously. Such highly distorted Fe-Co is predicted to reach a strong magnetocrystalline anisotropy which may compete with shape anisotropy. Probing this theoretical suggestion experimentally, epitaxial films are examined. A spontaneous strain up to \SI{5}{\%} lattice distortion is obtained for B contents up to \SI{4}{at\%}, which leads to uniaxial anisotropy constants exceeding \SI{0.5}{MJ/m$^3$}. However, a further addition of B results in a partial amorphization, which degrades both anisotropy and magnetization. 

\end{abstract}


\maketitle

\section{Introduction}

Fe-Co exhibits one of the highest magnetizations among all magnetic materials~\cite{Victora1984,James1999} and is thus very attractive for applications~\cite{Coey2001}. Fe-Co-B alloys are well known as soft magnetic materials due to the glass forming ability of boron~\cite{Ohandley1977,Kim2004,Yang2008,Asai2013}. In such amorphous materials the magnetocrystalline anisotropy is suppressed. In this study, we demonstrate that Fe-Co-B is also a promising hard magnetic material, when the amorphization is avoided and the B atoms induce a spontaneous strain in the crystalline lattice.

Within this introduction, we will first summarize criteria for soft and hard magnetic properties. We then use Density Functional Theory (DFT) to show that introducing B on interstitials sites in the Fe-Co lattice results in spontaneously strained phases with substantial magnetocrystalline anisotropy (MCA). In order to probe the limits of this approach experimentally, we study a series of epitaxial thin films. Results are discussed with respect to the concepts of epitaxial and spontaneous strain in Fe-Co for obtaining rare-earth free permanent magnets. In particular, we relate Fe-Co-B presented here to recently published Fe-Co-C films~\cite{Reichel2014}.

Since both, soft and hard magnets require a high magnetization, Fe-Co~\cite{Victora1984,James1999} is in focus of research. For soft magnetic materials, the magnetic anisotropy should be as low as possible, which is commonly achieved by reducing the grain size towards nanocrystalline or even amorphous materials~\cite{Asai2013,Kim2004,Minor2002}. Aiming at hard magnetic properties, a crystalline structure is essential as a magnetocrystalline anisotropy can favor particular directions of magnetization. The bcc crystal structure of binary Fe-Co, however, exhibits only a low cubic anisotropy~\cite{Shih1934}. Reducing the crystal symmetry by uniaxial strain is a route to achieve high magnetocrystalline anisotropy in Fe-Co as proposed in different theoretical studies~\cite{Burkert2004,Neise2011,Kota2012,Turek2012}. According to these studies, tetragonally strained Fe-Co is considered a possible rare-earth free permanent magnet material.
Due to its high magnetostriction~\cite{Hall1960,Pfeifer1980}, Fe-Co indeed appears to be more susceptible to strain compared to e.\,g.\,Fe-Ni, which makes Fe-Co less favorable regarding soft magnet applications. Though inverse magnetostriction only describes the influence of low strains, it allows a speculation that large strains are beneficial for a high magnetocrystalline anisotropy.

The tetragonal strain is commonly described by the lattice parameters of the strained axis $c$ and the compressed axes $a$, perpendicular to $c$. There are two main routes to strain the Fe-Co lattice experimentally. First, in coherent epitaxial growth, Fe-Co thin films adapt the in-plane lattice parameter $a$ from the substrate, which results in an induced strain ($c/a>1$) due to compressive stress within the film's plane. However, the driving forces for strain relaxation are too high to maintain tetragonal distortion in films thicker than 15 monolayers~\cite{Andersson2006,Winkelmann2006,Luo2007,Yildiz2008,Yildiz2009}, which is equivalent to \SI{2}{nm}. In situ studies~\cite{Reichel2014} revealed that binary Fe-Co films of \SI{4}{nm} thickness are again cubic with $c/a=1$. In order to stabilize the strain, calculations based on Density Functional Theory (DFT)~\cite{Delczeg2014} motivated an introduction of C atoms on interstitial sites in Fe-Co. Delczeg et al.~\cite{Delczeg2014} proposed that about \SI{6}{at\%} of these small atoms establish a $c/a$ ratio of 1.12. In experiments, however, the solubility on interstitial sites is much more limited. Using non-equilibrium preparation methods like Pulsed Laser Deposition (PLD), the limit of solubility, which is only about \SI{0.1}{at\%} for C or B in bcc Fe-Co in thermal equilibrium~\cite{Cameron1986}, can be shifted to significantly higher values. PLD prepared Fe-Co films containing \SI{2}{at\%} C exhibit such spontaneous strain with $c/a=1.03$ up to at least \SI{100}{nm} thickness as shown in our previous study~\cite{Reichel2014}. In contrast to the induced tetragonal strain at a coherent epitaxial interface, interstitials can result in a minimum of total energy at a certain lattice distortion. This second approach is thus not limited to ultrathin films and accordingly, the strain is also present in thicker films. In analogy to martensitic transformations, we assign this distortion as spontaneous strain, though a direct examination of an associated transformation is difficult, as common DFT only gives the ground state and in thin films, a transformation is often hindered by the rigid substrate~\cite{Buschbeck2009}. Nevertheless, the energy differences calculated between cubic and tetragonal state are within a factor of two comparable to thermal energies at room temperature and thus transformations are expected to occur in bulk materials.

Computational modeling can act as a powerful guide to experimental studies. Hence, we begin our work by utilizing a combination of the most relevant and highly accurate, spin-polarized DFT methods. In the first step, structural properties are modeled with low dopant concentrations being reproduced by using large supercells and alloying treated in the coherent potential approximation (CPA). The result of these structural studies are then used as input for the next step, where magnetic properties, including magnetic moments of both Fe and Co atoms on each inequivalent site, as well as MAE and orbital moments, are evaluated in fully relativistic calculations including the effect of the crucial spin-orbit coupling. 

In PLD, ions with high kinetic energy around \SI{100}{eV} are deposited~\cite{Faehler1996}. Thus, besides regular deposition, an implantation of the material within the topmost film monolayer takes place~\cite{Faehler1999} and allows for a supersaturation of Fe-Co films with C or B. However, this high energy impact also supports the formation of lattice defects. Supersaturated PLD prepared Fe-Co films with \SI{2}{at\%} C retain $c/a$ ratios of 1.03 up to high film thicknesses~\cite{Reichel2014}. In such spontaneously strained Fe-Co-C, magnetocrystalline anisotropy energies (MAE) of about \SI{0.4}{MJ/m^3} were observed independently from film thickness. Fe-Co with spontaneous strain should not have limitations in film thickness and is thus considered more promising than Fe-Co films with induced strain. 

An open question, though, is how the spontaneous strain and the MCA may be further increased. Since the solubility of C in Fe-Co is limited also in PLD prepared films~\cite{Reichel2014}, we focus on boron as atom of comparable size. The location of B in bcc Fe (and similar in bcc Fe-Co) is still under debate. Boron can substitute Fe in the ideal bcc positions \cite{Busby1953,Shevelev1958,Strocchi1967}, can occupy interstitial positions \cite{Hasiguti1954,Thomas1955,Tavadze1965,Wang1995}, and B solution in Fe can be stabilized by Ni, N and C~\cite{Hayasi1970,Lucci1971} both, interstitially and substitutionally~\cite{Fors2008}. In a more recent theoretical investigation~\cite{Baik2010}, it was shown that B prefers the substitutional chemical disorder in bulk Fe-B. However, this work did not take a local relaxation of the strain around interstitial B atoms into account. The authors further found that surfaces stabilize the occupation of octahedral interstitials. Near free surfaces, the geometrical pressure on B interstitials relaxes. Occupation of octahedral interstitials is followed by an increase of the B-Fe distances and thus a distortion of the bcc lattice. Baik et al.~\cite{Baik2010} thus give the motivation to further study the straining effect of B interstitials in Fe(-Co) thin films which are also dominated by free surfaces.

Considering the different interstitial sites in a bcc lattice, a tetragonal strain is exclusively expected, when the B atoms occupy octahedral interstitials. Substitutional arrangements or occupation of tetrahedral interstitials would not change $c/a$, but only affect the unit cell volume. Octahedral interstitials as defined in Ref.~\cite{Cahn1996} exist along all three spatial directions (see supplementary~\cite{supp}). There are six octahedral interstitial sites in a bcc lattice, indicating a high theoretical solubility, when the atomic radius of the interstitial atom is sufficiently low and chemical conditions are beneficial. If such an interstitial site is occupied by a small atom like B, the atoms of the apexes of the octahedron displace along its axis, i.\,e. uniaxially. A lattice strain by means of $c/a>1$ is thus only possible, when the octahedral interstitials along the $c$ axis as e.\,g. (0;0;1/2) within the bcc unit cell are preferentially occupied. The question, how an additional occupation of the other octahedral institials affects the structural properties and the total energy will be discussed based on DFT calculations.

Besides the discussed lattice preferences of boron, amorphization of the Fe-Co lattice is expected to begin at a certain B content. Depending on the film preparation conditions, amorphous Fe-Co-B phases were reported at B concentrations of e.\,g.\,7.5~\cite{Yang2008}, 15~\cite{Asai2013} or \SI{22}{at\%}~\cite{Kim2004}. In order to establish a high magnetocrystalline anisotropy in Fe-Co-B, amorphization has to be avoided. Our study thus presents theoretical calculations of the preference of B atoms to occupy the same type of octahedral interstitials due to lower energy of such configuration. Then it focuses on the MCA evaluation of ideal Fe-Co-B crystals assuming all B impurities occupying the octahedral interstitials perpendicular to the substrate. We then introduce the properties of PLD prepared epitaxial Fe-Co-B films, taking the different possibilities how B may affect the Fe-Co lattice into consideration.

\section{Methods}
\subsection{Computational methods}

First principles electronic structure calculations within the generalized gradient approximation (GGA)~\cite{Perdew1996} were used to identify stable or metastable body centered tetragonal (bct) structures for B doped Fe-Co alloys.

In the first step, different B concentrations were modeled by three Fe$_y$B supercells, Fe$_8$B, Fe$_{16}$B and Fe$_{24}$B with $1 \times 2 \times 2$, $2 \times 2 \times 2$ and $2 \times 2 \times 3$ supercells, respectively. Fe atoms occupy the ideal bcc positions, while B atoms are placed in the octahedral positions, which are the most stable interstitial positions~\cite{Baik2010}. B dopants in the octahedral positions reproduce the upper limit for tetragonal distortion. Substitutional B dopants were not considered as they do not contribute to the formation of distortion. The conjugate-gradient algorithm as implemented in the Vienna Ab Initio Simulation Package (VASP)~\cite{Kresse1996I,Kresse1996II,Blochl1994,Kresse1999} was used to fully relax these structures. The $\mathbf{k}$-point mesh was set to $16 \times 16 \times 8$, $8 \times 8 \times 8$, $8 \times 8 \times 6$ for Fe$_8$B, Fe$_{16}$B and Fe$_{24}$B, respectively, within the Monkhorst-Pack scheme~\cite{Monkhorst1976}. The plane-wave cut off and the energy convergence criterion of the scalar relativistic calculations were set to \SI{500}{eV} and 10$^{-7}$\,eV, respectively.

In the second step, the alloying effect on the equilibrium parameters ($c/a$)$_{\rm eq}$-ratio and volume $V_{\rm eq}$ of these relaxed structures was studied using the Exact Muffin-Tin Orbitals - Full Charge Density (EMTO-FCD) method~\cite{Andersen1994,Andersen1998,Vitos2000,Vitos2001, Vitos2001a,Vitos2007,Vitos1997,Kollar2000}. The accuracy of the EMTO method is controlled by the optimized overlapping muffin-tin potential~\cite{Vitos2001,Vitos2007,Zwierzycki2009}. The muffin-tin potential optimization procedure is described in details in Ref.~\cite{Al-Zoubi2012}. The potential parameter $\eta$ was chosen to match the equilibrium parameters obtained by EMTO to those obtained by the VASP calculations. The potential parameter $\eta$ is 0.75, 0.85 and 0.9 for Fe$_8$B, Fe$_{16}$B and Fe$_{24}$B, respectively. The chemical disorder for (Fe$_{1-x}$Co$_x$)$_y$B alloys was treated via the Coherent Potential Approximation (CPA)~\cite{Soven1967,Gyorffy1972}. The one-electron equations were solved within the scalar-relativistic and soft-core approximations. The Green's functions were calculated for 18 complex energy points distributed exponentially on a semi-circular contour with radius of \SI{1.2}{Ry}. The 3$d$ and 4$s$ states of Fe, Co, and the 2$s$ and 2$p$ states of B were treated as valence electrons. $s$, $p$, $d$, $f$ orbitals were included in the muffin-tin basis set ($l_\text{max}=3$). The one center expansion of the full-charge density was truncated at $l_\text{max}^h=8$. Between 40 and 500 uniformly distributed $\mathbf{k}$-points were used in the irreducible wedge of the Brillouin zone. The electrostatic correction to the single-site CPA was described using the screened impurity model~\cite{Korzhavyi1995,Ruban2002} with a screening parameter of 0.6. Other screening parameters~\cite{Rahaman2011} have been tested for the smallest system and it turned out that their effect on the equilibrium parameters is less than \SI{1}{\%}.

The magnetic properties, including magnetic moments and the magnetocrystalline anisotropy energy (MAE), of the various (Fe$_{1-x}$Co$_x$)$_y$B systems were evaluated using the spin-polarized relativistic KKR (SPR-KKR)~\cite{Ebert2011,Ebert2012} method in the atomic spheres approximation (ASA), in a similar manner as was done in Ref.~\cite{Delczeg2014}. Calculations were performed using GGA~\cite{Perdew1996} for the exchange-correlation potential and at least several thousand $\mathbf{k}$-points (depending on system size) were used for numerical integration over the Brillouin zone in order to obtain well converged values of MAE. Alloying was treated with the CPA~\cite{Soven1967} and the MAE was obtained by total energy difference for two different magnetization directions, i.\,e. $\text{MAE} = E_\text{total}(\hat{m}\parallel100) - E_\text{total}(\hat{m}\parallel001)$.

All structures from subsection~\ref{preferential} were fully relaxed, $c/a$ ratios and volumes were optimized together with atomic relaxation for every $c/a$ ratio and volume.
Those calculations were performed with the full potential linearized augmented plane wave method (FP-LAPW) implemented in the WIEN2k code~\cite{Blaha2001}.
For the exchange-correlation potential the GGA form~\cite{Perdew1996} was used. 
Calculations were performed with a plane wave cut-off parameter $RK_\text{max}=6.5$, total energy convergence criterion $10^{-6}$~Ry, with $4 \times 4 \times 4$ \textbf{k}-points and with radii of the atomic spheres 1.96\,$a_0$ for Fe/FeCo and 1.45\,$a_0$ for B, where $a_0$ is the Bohr radius.
The Virtual Crystal Approximation (VCA) was used to study different Fe$_{0.4}$Co$_{0.6}$-B compositions, following the same procedure as described in Ref.~\cite{Delczeg2014}.
All parameters were carefully tested to provide well converged values.

\subsection{Experimental}

The Fe-Co-B samples were prepared as thin films performing Pulsed Laser Deposition (PLD) in ultra-high vacuum ($5\times 10^{-9}$\,mbar) at room temperature. The used KrF excimer laser (Coherent LPXpro 305) operates with a wave length of \SI{248}{nm} and a pulse length of \SI{25}{ns}. On MgO(100) single crystal substrates, \SI{3}{nm} Cr seed layers and \SI{30}{nm} Au$_{0.55}$Cu$_{0.45}$ buffer layers were deposited prior to the deposition of the \SI{20}{nm} thick Fe-Co-B films. The Au-Cu and the Fe-Co-B layers were prepared in pseudo-co-deposition, i.\,e. by repetitive changing of the targets during PLD to achieve the aimed compositions. Elemental (Au, Cu, Fe and Co) and a FeB composite target were used therefore. The deposition rates were measured prior the preparation with a quartz crystal rate monitor.

Energy dispersive x-ray spectroscopy (EDX) measurements on a Bruker EDX in a JEOL JSM6510-NX electron microscope confirmed the intended film compositions. The boron content in the Fe-Co-B films was measured exploiting Auger Electron Spectroscopy (AES) on a JEOL JAMP-9500F Field Emission Auger Microprobe device. The atomic concentrations~\cite{Bcontent} were calculated using standard single element sensitivity factors from mean values of sputter depth profiles measurements carried out with \SI{1}{keV} Ar$^+$ ions. Atomic force microscopy (AFM) measurements on an Asylum Research Cypher AFM were performed to study the surface morphologies and to confirm a high flatness of the film surfaces. A Bruker D8 Advance diffractometer operating with CoK$\alpha$ radiation was used for X-ray diffraction (XRD) in Bragg-Brentano geometry. The Fe-Co-B lattice strains were derived from pole figure measurements, which had been carried out on an X'pert four circle goniometer (CuK$\alpha$ radiation). Transmission electron microscopy (TEM) studies were performed on a Titan$^3$ 80-300 microscope, which was equipped with a C$_\text{S}$ corrector and a Schottky field emission electron source. The lamella was cut using focused ion beam on a FEI microscope (Helios NanoLab 600i). For magnetic hysteresis measurements at \SI{300}{K}, a vibrational sample magnetometer (VSM) mounted in a Quantum Design Physical Property Measurement System was used.

\section{Results and discussion of DFT studies}

\subsection{Structure optimization}

\begin{figure*}[th]
\begin{center}
 \includegraphics[width=15cm,trim= 0mm 120mm 0mm 5mm, clip=true]{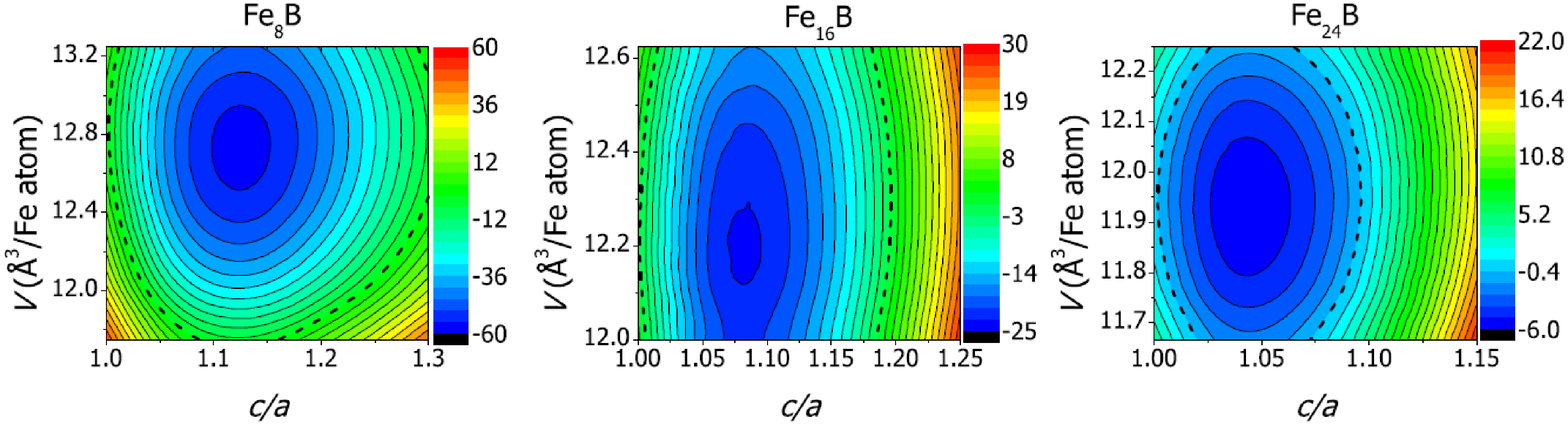}
\caption{(Color online) Contour plots of total energy (meV/Fe atom) as a function of $c/a$-ratio and unit volume $V$, for Fe$_{8}$B, Fe$_{16}$B and Fe$_{24}$B, respectively. The zero reference values for the energies are $E$($c/a$=1, $V_{\rm {eq}}^{c/a \rm {=1}}$). The zero energy isolines are denoted by dashed lines. Total energy and volume are normalized per Fe atom.}
\label{fig_vasp}
\end{center}
\end{figure*}

In order to prepare crystal structures of chemically disordered (Fe$_x$Co$_{1-x}$)$_y$B, 
the simpler Fe$_y$B systems have to be considered as an initial step.
Fe$_y$B structures are modeled with B atoms placed into octahedral interstitial positions along the $c$ axis.
This configuration determines the upper limit of the tetragonal distortion. 
Internal relaxation of Fe atoms around the B interstitials leads to the elongation of $c$ and shrinking of $a$ lattice parameter.
In other words, the $c/a$ ratio is now higher than 1, where 1 corresponds to a cubic lattice.
The total energies $E$ as a function of $c/a$ ratio and volume $V$ for three Fe$_y$B systems are computed by VASP and their plots are presented in Figure~\ref{fig_vasp}.

At each $c/a$ ratio, the equilibrium volume ($V_{\rm eq}$) is obtained by using a Morse-type equation of state as minimum of $E(V)$ dependency. 
Next, the equilibrium ratio $(c/a)_{\rm eq}$ is attained by fitting $E(V_{\rm eq})$ with a second order polynomial. 
Results of $(c/a)_{\rm eq}$ and $V_{\rm eq}$ are listed in Table~\ref{tab1}.
It is observed that both the $c/a$ ratio and volume increase with B content, as expected.

\begin{table}[ht!]
\caption{Equilibrium $(c/a)_{\rm eq}$ ratios and volumes ($V_{\rm eq}$) of Fe$_y$B systems obtained by EMTO and VASP.}
\centering
\begin{tabular}{l|c|ccc|cc}
  \hline\hline
Alloy      &at\% B&\multicolumn{3}{c|}{$(c/a)_\text{eq}$} & \multicolumn{2}{c}{$V_{\rm eq}$(\AA{}$^3$/Fe atom)} \\
           &       &   VASP & EMTO  & $\Delta$&  VASP & EMTO\\
  \hline
Fe$_8$B    & 11.1  &  1.126 & 1.132 & -0.006  & 12.73 &12.88\\
Fe$_{16}$B & 5.9   &  1.089 & 1.084 &  0.005  & 12.21 &12.25\\
Fe$_{24}$B & 4.0   &  1.050 & 1.044 &  0.006  & 11.93 &11.95\\
  \hline\hline
\end{tabular}\label{tab1}
\end{table}

\begin{figure*}[th]
\begin{center}
 \includegraphics[width=0.28\columnwidth]{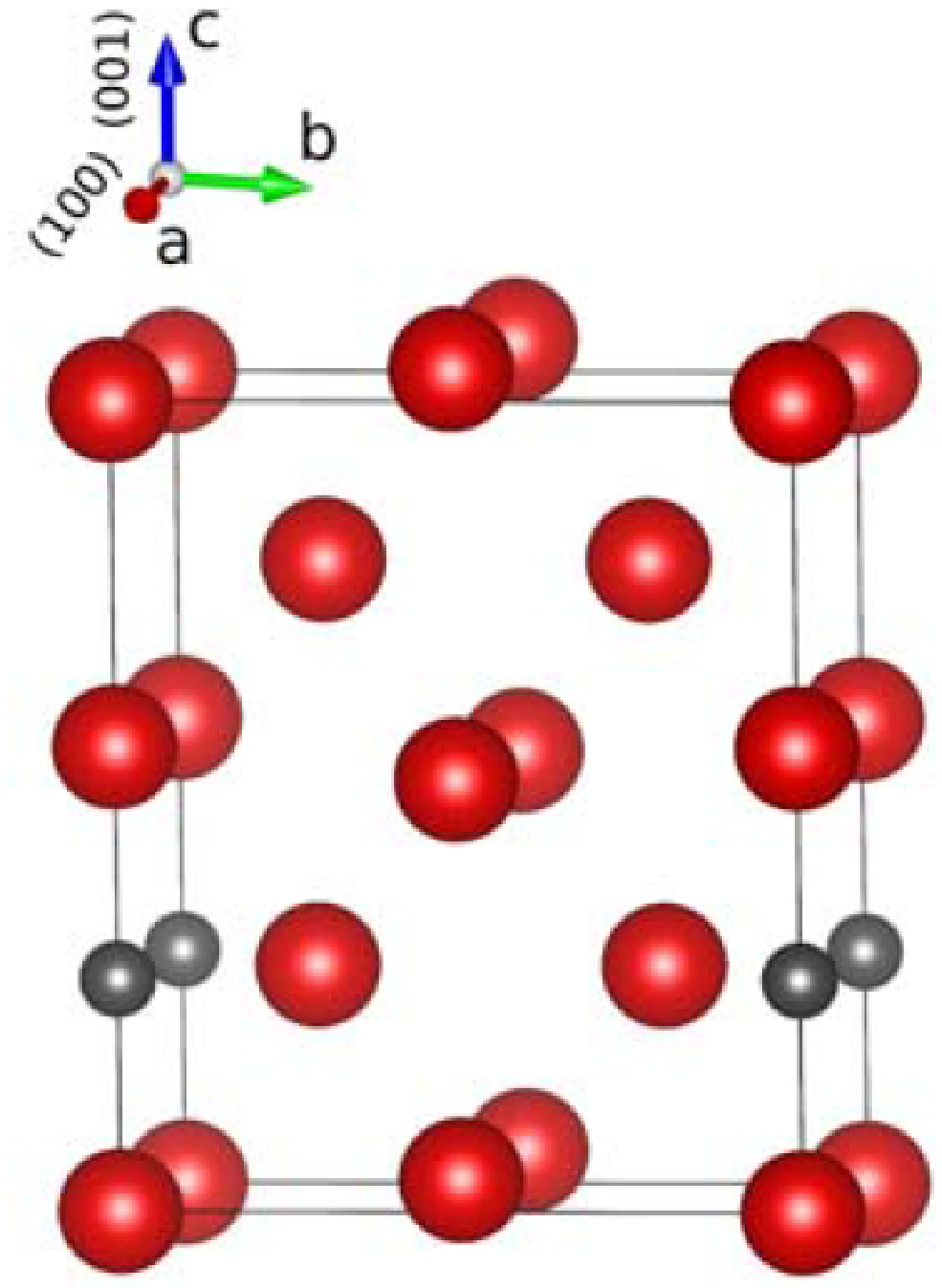}
 \includegraphics[width=0.3\columnwidth]{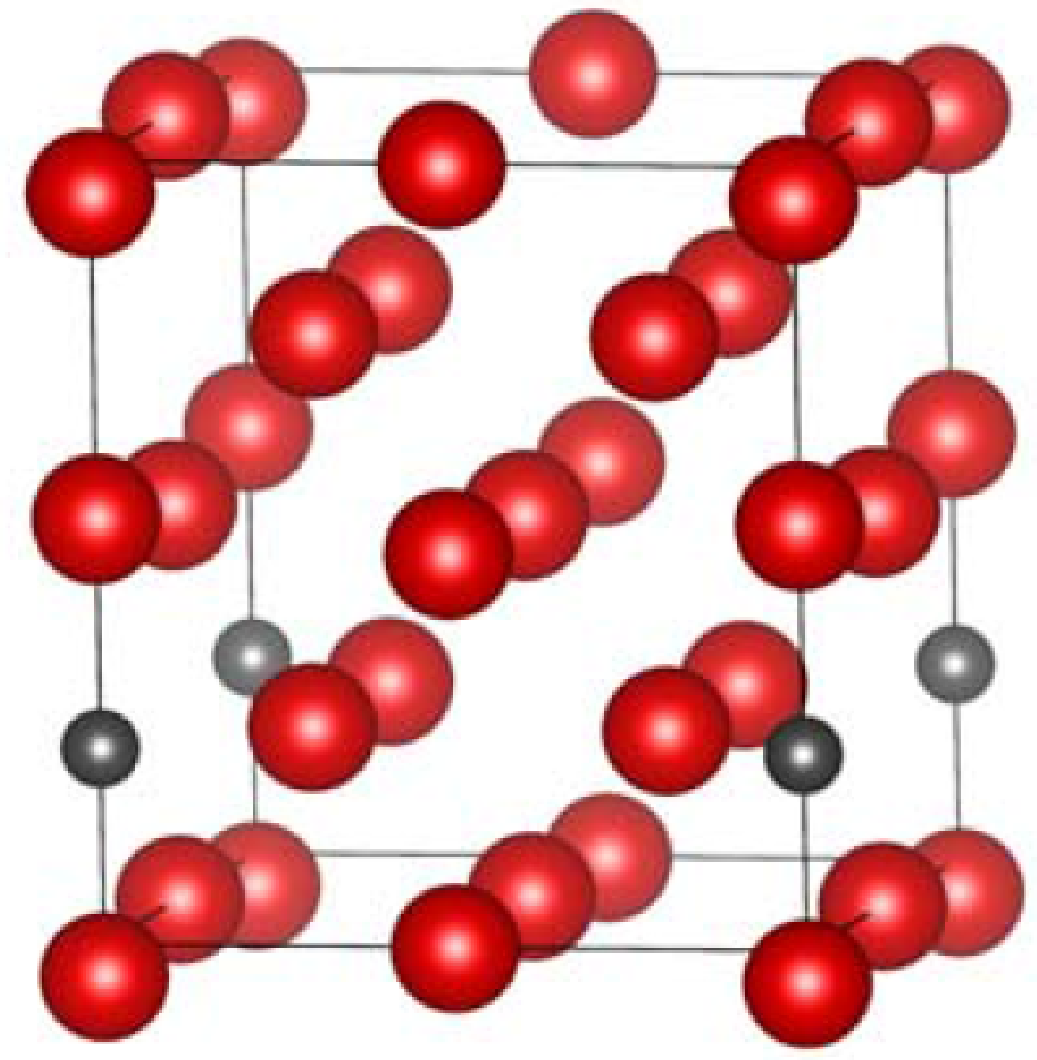}
 \includegraphics[width=0.3\columnwidth]{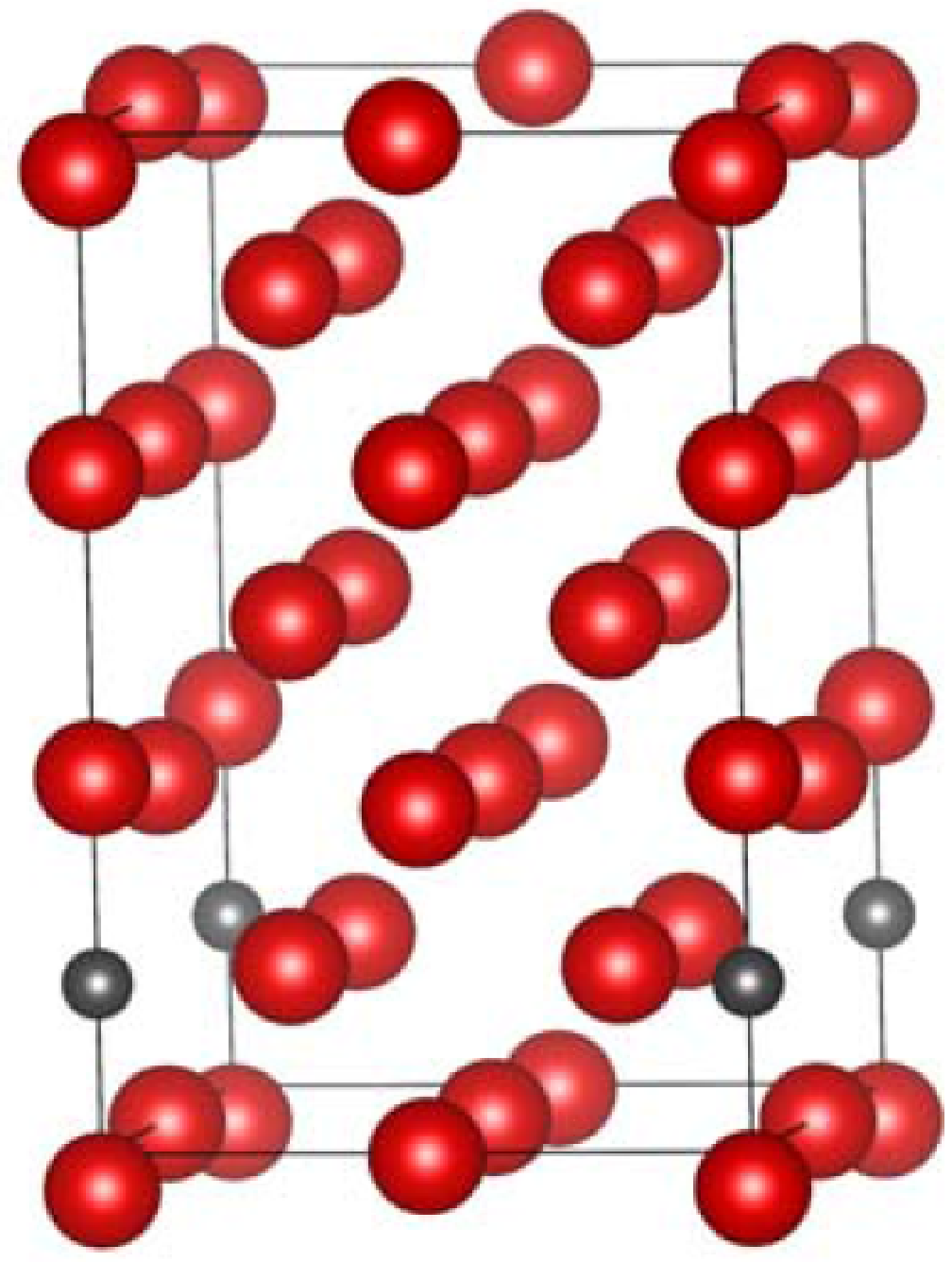}
\caption{(Color online) Crystallographic models of (Fe$_{x}$Co$_{1-x}$)$_{8}$B, (Fe$_{x}$Co$_{1-x}$)$_{16}$B and (Fe$_{x}$Co$_{1-x}$)$_{24}$B, respectively. The bigger red spheres represent Fe/Co atoms while smaller black spheres denote B atoms.}
\label{fig_structures}
\end{center}
\end{figure*}

Three types of (Fe$_{x}$Co$_{1-x}$)$_{y}$B supercells corresponding to different B concentrations have been considered, i.\,e. (Fe$_{x}$Co$_{1-x}$)$_{8}$B, (Fe$_{x}$Co$_{1-x}$)$_{16}$B and (Fe$_{x}$Co$_{1-x}$)$_{24}$B, see Figure~\ref{fig_structures}.
Initially, (Fe$_{x}$Co$_{1-x}$)$_{8}$B and (Fe$_{x}$Co$_{1-x}$)$_{16}$B were taken into account as reasonable crossing point between computational effort and expected experimental B solubility.
Preliminary results suggested that higher B concentration gives higher $c/a$ ratio and MAE.
Unfortunately, neither ~\SI{11.1}{at\%} B ((Fe$_{x}$Co$_{1-x}$)$_{8}$B) nor \SI{5.9}{at\%} B ((Fe$_{x}$Co$_{1-x}$)$_{16}$B) could be obtained experimentally, and what was shown instead, maximum B solubility producing maximum tetragonal strain is about \SI{4}{at\%}.
At that point of this study the (Fe$_{x}$Co$_{1-x}$)$_{24}$B structures (\SI{4}{at\%} B) were introduced and calculated.

The effect of substituting Fe by Co on the structural parameters of Fe$_{y}$B has been taken into account by using the CPA approximation implemented in the EMTO method.  
The equilibrium ratios $(c/a)_{\rm eq}$ and the lattice parameter $a_{\rm eq}$ have been evaluated for the rigid supercells taken from the preceding VASP calculations. 
Thus, it is important to assess the accuracy of the EMTO method compared to VASP for the pure Fe$_y$B.
For each $c/a$ ratio, a fully relaxed structure is generated by VASP at $V_{\rm eq}$ and used as input for the EMTO.
Total energies are then calculated for each $c/a$ ratio and five different $V's$ while keeping the internal parameters fixed.
Results obtained by EMTO, using the above mentioned fitting procedure for VASP, are listed in Table~\ref{tab1}. 
The error bar of the $(c/a)_{\rm eq}$ ratio is defined by $\Delta$ = $(c/a)^{\mathrm{VASP}}_{\rm eq}$ - $(c/a)^{\mathrm{EMTO}}_{\rm eq}$, with  
$c/a$ ratios obtained by VASP and EMTO.
Good agreement between these two theoretical methods, regarding both $(c/a)_{\rm eq}$ and $V_{\rm eq}$, is observed.

After validation of the EMTO method, the desired CPA structures of (Fe$_{x}$Co$_{1-x}$)$_{y}$B have been evaluated.
Results are presented in Table~\ref{MAEtable}.
It is observed that an increase of Co concentration in (Fe$_{x}$Co$_{1-x}$)$_{y}$B enhances the tetragonal distortion $(c/a)_{\rm eq}$.
At the same time the $a_{\rm eq}$ and $V_{\rm eq}$ decrease.
Decrease in  $V_{\rm eq}$ is ascribed to the lower atomic volume of Co compared to Fe. 

Note, that the tetragonal distortion predicted for B-doped Fe-Co alloys is larger than for C-doped alloys~\cite{Delczeg2014}.
This can be understood by the fact that the B atomic radius is bigger than the C atomic radius. 
By adding \SI{4}{at\%} of B a distortion around 1.05--1.07 should be achieved, while for C-doping it is merely about 1.03--1.04. 
This comparison suggests that B-doping could be a better way to achieve tetragonally distorted Fe-Co alloys.
%

Note further, that with B-doping, the tetragonal distortion can be obtained for a broader range of Co concentrations, compared to C-doped counterparts. This suggests a possibility to prepare tetragonal Fe-Co alloys with lower Co contents. In order to find out whether this brings any advantages, we discuss their magnetic characteristics in the following section.

\begin{table*}
    \caption{Summary of calculated structural (EMTO) and magnetic (SPR-KKR) characteristics of (Fe$_{x}$Co$_{1-x}$)$_{y}$B systems.}
	\begin{tabular*}{\textwidth}{@{\extracolsep{\fill} } c c c c c c c c c r c }
	    \hline\hline
	    Composition & $(c/a)_{\rm eq}$ & $a_{\rm eq}$ & $V_{\rm eq}$ & $m_S$ & $m_L$ & $\mu_0 M_s$ & MAE & MAE \\
	                &                  & (\AA{})      & ($\frac{\AA{}^3}{\text{Fe atom}}$) & ($\frac{\mu_{\text{B}}}{\text{atom}}$) & ($\frac{\mu_{\text{B}}}{\text{atom}}$) & (T) & ($\frac{\mu\text{eV}}{\text{atom}}$) &  ($\frac{\text{MJ}}{\text{m}^3}$) \\
	    \hline
	    (Fe$_{0.5}$Co$_{0.5}$)$_{8}$B    & 1.247 & 2.716 & 12.49 & 1.71 & 0.066 & 1.87 & 118 & 1.51 \\ \hline
	    (Fe$_{0.50}$Co$_{0.50}$)$_{16}$B & 1.086 & 2.809 & 12.04 & 1.97 & 0.068 & 2.09 &  46 & 0.62 \\
	    (Fe$_{0.45}$Co$_{0.55}$)$_{16}$B & 1.091 & 2.802 & 12.00 & 1.91 & 0.068 & 2.04 &  51 & 0.69 \\
	    (Fe$_{0.40}$Co$_{0.60}$)$_{16}$B & 1.103 & 2.788 & 11.95 & 1.86 & 0.069 & 2.00 &  52 & 0.69 \\
            (Fe$_{0.35}$Co$_{0.65}$)$_{16}$B & 1.116 & 2.774 & 11.91 & 1.82 & 0.069 & 1.96 &  43 & 0.58 \\ \hline
	    (Fe$_{0.65}$Co$_{0.35}$)$_{24}$B & 1.049 & 2.835 & 11.95 & 2.16 & 0.064 & 2.26 &  23 & 0.31 \\
	    (Fe$_{0.60}$Co$_{0.40}$)$_{24}$B & 1.051 & 2.832 & 11.94 & 2.12 & 0.066 & 2.22 &  30 & 0.40 \\
            (Fe$_{0.55}$Co$_{0.45}$)$_{24}$B & 1.053 & 2.827 & 11.90 & 2.08 & 0.067 & 2.19 &  30 & 0.40 \\
	    (Fe$_{0.50}$Co$_{0.50}$)$_{24}$B & 1.056 & 2.822 & 11.87 & 2.04 & 0.068 & 2.15 &  28 & 0.38 \\
            (Fe$_{0.45}$Co$_{0.55}$)$_{24}$B & 1.059 & 2.816 & 11.82 & 1.99 & 0.069 & 2.12 &  27 & 0.37 \\
	    (Fe$_{0.40}$Co$_{0.60}$)$_{24}$B & 1.063 & 2.810 & 11.79 & 1.95 & 0.070 & 2.08 &  29 & 0.40 \\
            (Fe$_{0.35}$Co$_{0.65}$)$_{24}$B & 1.068 & 2.803 & 11.76 & 1.91 & 0.070 & 2.04 &  35 & 0.48 \\ \hline
 \hline
	\end{tabular*}
    \label{MAEtable}
\end{table*}

\subsection{Preferential orientation of octahedral interstitials}
\label{preferential}

For a permanent magnet it is decisive that the easy axis of all unit cells is aligned along one particular direction, in which the magnet will be magnetized. The different orientations of octahedral interstitials~\cite{Cahn1996}, which could be occupied by the B atoms, thus have to be considered. Consequences of a random distribution have to be discussed as well. The occupation of tetrahedral sites, which just result in a volume change, are not considered in the following since they are energetically unfavorable. Baik et al.~\cite{Baik2010} computed that the energy difference between octahedral and tetrahedral sites amounts to \SI{0.70}{eV} for a cubic Fe-B supercell, without taking into account an optimization of lattice parameters. The corresponding energy difference calculated by us is \SI{0.77}{eV/B~atom}, based on bcc Fe 2x2x2 supercells (Fe$_{16}$B) with geometry optimization. This means, that B atoms favor the octahedral sites even stronger than Baik et al. showed. A similar set of calculations for the alloy (Fe$_{0.4}$Co$_{0.6}$)$_{16}$B, treated within the VCA, gives the energy difference \SI{0.39}{eV/B~atom}, which also suggests to consider only the octahedral interstitials in the subsequent models.

As our calculations reveal a tetragonal distortion with uniaxial anisotropy along the $c$ axis, independent of the B content, we can take any tetragonal building block for the following considerations, which are borrowed from martensite theory as this can be extrapolated towards the unit cell level~\cite{Kaufmann2010,Kauffmann2011}. We consider martensite theory as reasonable since the diffusion of B as interstitial~\cite{Wang1995} is much faster in contrast to Fe (or Co) atoms~\cite{Buffington1961}. In this picture, the orientation of the tetragonal distortion of the Fe-Co "martensite" changes, when B moves from one octahedral site to another. 

The connection of different tetragonal building blocks is possible by twin boundaries, but they require additional twin boundary energy. A twin boundary is connected with a 90$^\circ$ domain wall in these uniaxial ferromagnets, which further requires magnetic exchange energy. In this simplified picture, both, elastic and magnetic contributions make a different orientation of neighboring tetragonal building blocks energetically unfavorable. Twin boundaries must be introduced, however, to minimize elastic stress energy and magnetostatic energy, and in bulk samples all three orientations are thus equivalent and should occur together. 

\begin{figure}[ht]
\centering
\includegraphics[width=0.49\columnwidth]{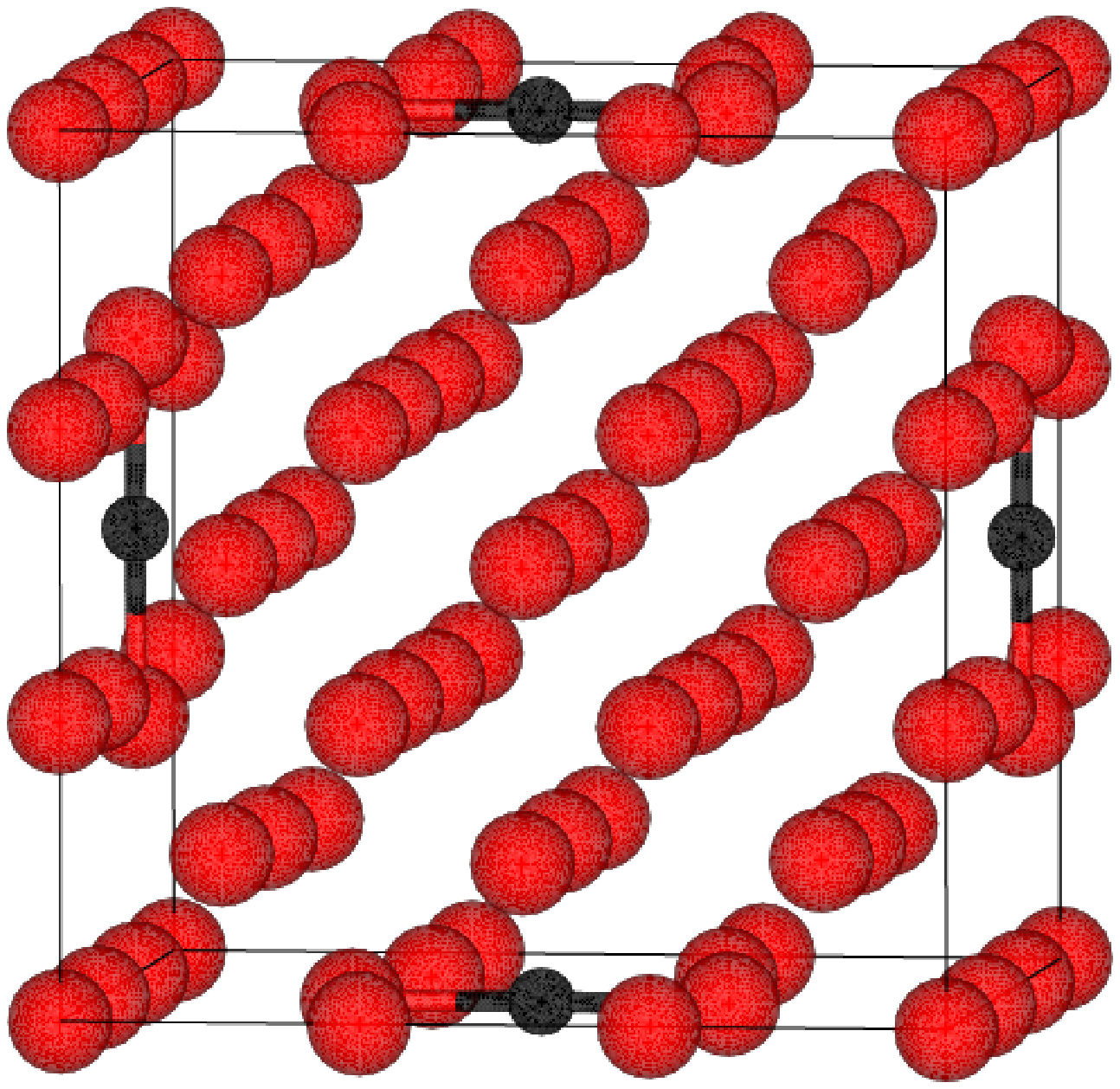}
\includegraphics[width=0.49\columnwidth]{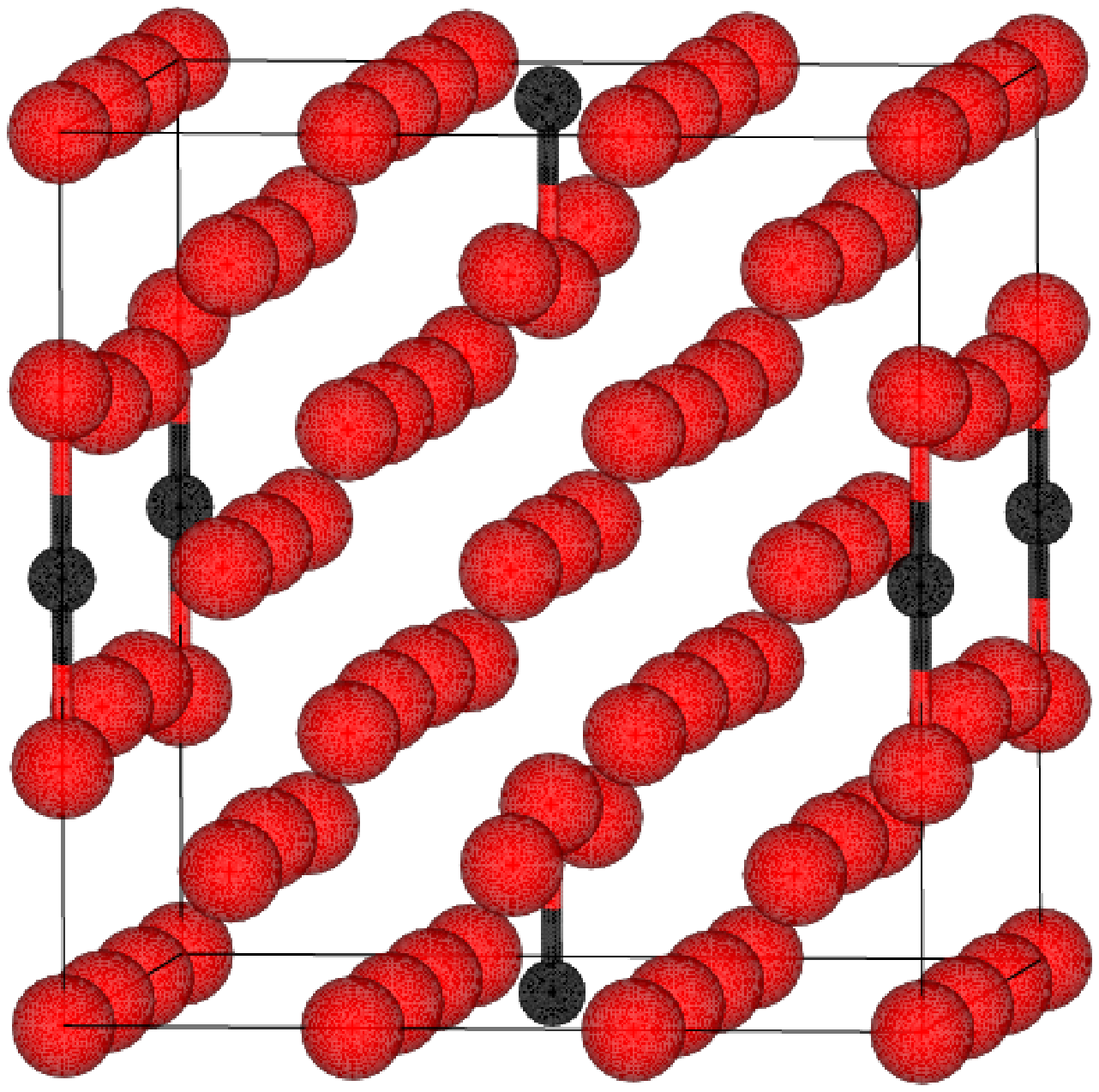}
a) Fe$_{54}$B$_{2}$ orthogonal  \hspace{10mm} b) Fe$_{54}$B$_{2}$ uniaxial
\caption{3x3x3 bcc Fe supercells with two B atoms in octahedral interstitial positions (Fe$_{54}$B$_{2}$) as a) orthogonal and b) uniaxial configuration of interstitial axes.}
\label{fig:fe54b2_structs}
\end{figure}

In order to support this picture on the atomic scale, we performed DFT calculations of a configuration with orthogonal B interstitials and compare it with a uniaxial orientation of B interstitials. For this comparison, one has to consider at least two B atoms per unit cell, whereby the Fe/FeCo supercell has to be enlarged in order to keep low B concentration. In the first step, two Fe based models are considered, both with two B atoms in octahedral interstitial positions in bcc Fe 3x3x3 supercells (Fe$_{54}$B$_{2}$), see Fig.~\ref{fig:fe54b2_structs}. 
In the orthogonal model (Fig.\,\ref{fig:fe54b2_structs}(a)), the positive strains are generated along two orthogonal axes, which leads to a $c/a$ ratio below 1. 
In the second model (Fig.\,\ref{fig:fe54b2_structs}(b)), the octahedral interstitial axes are oriented parallel, which forms strain exclusively along the $c$ axis and leads to $c/a$ ratio above 1.
Calculations show that the uniaxial configuration is favored with a difference of total energy equal to \SI{129}{meV/B~atom}.
In order to relate the theoretical result to the experimental Fe-Co-B system one has to go beyond the pure Fe model and, as we did in the previous section, consider the influence of alloying with Co. For this reason, in the second step, the two Fe$_{54}$B$_{2}$ models described above have been recalculated for a selected Fe$_{0.4}$Co$_{0.6}$ concentration within the VCA. Results show once again that the uniaxial configuration is more stable, with slightly lower total energy difference of \SI{126}{meV/B~atom}. CPA calculations are prohibitively computationally expensive for the EMTO method due to a large number of inequivalent atoms in the unit cells. Some crystallographic data of the optimized structures are collected in Table~\ref{tab:fe54b2_crystal_data}. An interesting result is that the supercell volumes, which for orthogonal and uniaxial configurations were optimized independently, have the same values up to 5 significant digits, which indicates the high accuracy of these results.

\begin{table}[ht]
\caption{\label{tab:fe54b2_crystal_data} 
Crystallographic data of optimized Fe$_{54}$B$_{2}$ and (Fe$_{0.4}$Co$_{0.6}$)$_{54}$B$_{2}$ structures and differences of total energies between  orthogonal and uniaxial configurations of B interstitials.} 
\begin{tabular}{c|ccc|ccc|c}
\hline \hline
			&\multicolumn{3}{c|}{orthogonal} & \multicolumn{3}{c|}{uniaxial} & \\
					
			& $V$ 	 & $a$ 	& $c/a$	& $V$	  & $a$    & $c/a$	& $\Delta{}E$  \\
 			&[\AA{}$^3$]&[\AA{}]& 	&[\AA{}$^3$]&[\AA{}]&  	& $\frac{\text{meV}}{\text{B atom}}$ \\
\hline
Fe$_{54}$B$_{2}$			& 635.74& 8.652 & 0.982	& 635.74  & 8.492& 1.038 & 129\\
(Fe$_{0.4}$Co$_{0.6}$)$_{54}$B$_{2}$    & 625.31& 8.636 & 0.971	& 625.31  & 8.454& 1.035 & 126\\
\hline\hline
\end{tabular}
\end{table}

With three B atoms per 3x3x3 bcc Fe supercell it is possible to construct a system with octahedral interstitials in each of the three spatial directions, resulting in a cubic structure. Such a Fe$_{54}$B$_{3}$ structure with a simulated random occupation of the octahedral interstitials, while keeping their mutual distances as large as possible within the supercell, has been carefully optimized together with its uniaxial Fe$_{54}$B$_{3}$ counterpart. The total energy difference between these structures is \SI{49}{meV/B~atom} and the preferred configuration of boron is again the uniaxial one. After dividing the latter value by Boltzmann constant one obtains \SI{570}{K/B~atom}, the same order of magnitude as room temperature. This implies, that a small fraction of B atoms will occupy positions not yielding a tetragonal strain. Nevertheless, calculations of the presented ideal cases, which may not compare the energies of all theoretical possibilities, but compare the most extreme scenarios, suggest that an occupation of neighboring orthogonal octahedral sites with varying orientation is energetically unfavorable.


From an experimental point of view, the preferential site occupation is tedious to be measured directly due to the much lower atomic number of B compared to Fe-Co. However, the reduced symmetry of thin films compared to bulk may result in a favorable orientation, which could be induced by coherent epitaxial growth, followed by a relaxation towards the spontaneously strained state as shown for Fe-Co-C films~\cite{Reichel2014}. A preferential lattice orientation allows using integral methods like texture measurements to probe the present tetragonal distortion and magnetization measurements to determine the magnetocrystalline anisotropy. An agreement of these global measurements with the local DFT calculations would also confirm the unique alignment of all easy axes, which is beneficial for permanent magnet applications. 

In other words, when comparing DFT calculations with experiments, one has to consider the different length scales. In DFT calculations, the preferential $c$ axis is the result of the spontaneous strain at the atomic scale, while in thin film experiments, the out-of-plane orientation is given and may act on the whole sample. The expected preferential alignment of the strained $c$ axis along the out-of-plane orientation thus has to be probed by the following experiments.

\subsection{Saturation magnetization}

For the considered (Fe$_{x}$Co$_{1-x}$)$_{y}$B systems, the influence of B on magnetic moments is regarded first.
The average magnetic moments per atom, obtained by SPR-KKR calculations, are listed in Table~\ref{MAEtable}.
It is observed that the average magnetic moment decreases by increasing the B content.
This can be simply explained by addition of a non-magnetic B component itself, 
and furthermore by the reduction of the magnetic moments on Fe/Co atoms around the B interstitial.
In order to look closer at the latter effect an insight into atom and site specific magnetic moments is necessary.
Internal relaxation changes the distances between Fe/Co atoms, especially around B interstitials.
These reconfigurations affect the exchange coupling between Fe/Co atoms and thus affects magnetic moments.

\begin{table}[ht!]
\caption{Magnetic moments of (Fe$_{0.5}$Co$_{0.5}$)$_{y}$B systems ($\mu_{\text{B}}$/atom) obtained by SPR-KKR. Moments on Fe and Co atoms in the nearest neighborhood of B octahedral interstitial and as far as possible from B.}
\centering
\begin{tabular}{c|c|cc|cc}
  \hline\hline
Alloy                          &       &\multicolumn{2}{c|}{close to B} & \multicolumn{2}{c}{far from B} \\
                               &at\% B&  $m_\text{Fe}$ & $m_\text{Co}$  & $m_\text{Fe}$&  $m_\text{Co}$\\
  \hline
(Fe$_{0.5}$Co$_{0.5}$)$_{8}$B  & 11.1  &  1.74 & 0.98 & 2.89 & 1.95 \\
(Fe$_{0.5}$Co$_{0.5}$)$_{16}$B & 5.9   &  1.52 & 0.82 & 2.85 & 1.93 \\
(Fe$_{0.5}$Co$_{0.5}$)$_{24}$B & 4.0   &  1.52 & 0.85 & 2.76 & 1.91 \\
  \hline\hline
\end{tabular}\label{moments}
\end{table}

In Table~\ref{moments} the local magnetic moments for selected (Fe$_{0.5}$Co$_{0.5}$)$_{y}$B systems are presented. The Fe/Co atoms that are the closest neighbors to the B atoms have the magnetic moments most reduced compared to the moments on Fe/Co atoms far from B. The latter values alter rarely, indicating that the Fe/Co atoms far from B have bulk-like characteristics already for (Fe$_{0.5}$Co$_{0.5}$)$_{8}$B. This suggests that the influence of a B impurity is relatively short-ranged.

An increase of Co content leads to a decrease of the total magnetic moment 
for the considered Co concentrations in the (Fe$_{x}$Co$_{1-x}$)$_{24}$B system (see Table \ref{MAEtable}). This negative dependence is not necessarily true beyond the studied range, since for bct Fe/Co alloys a magnetization maximum is observed for lower Co concentrations~\cite{Victora1984,James1999}. In our case the drop of magnetic moment is accompanied by an increase of $(c/a)_{\rm eq}$.

\subsection{Magnetocrystalline anisotropy energy}

Table~\ref{MAEtable} summarizes the MAEs of the (Fe$_{x}$Co$_{1-x}$)$_y$B systems, as obtained by SPR-KKR calculations. Trends are in many aspects similar as those obtained for interstitial carbon~\cite{Delczeg2014}. In particular, the tetragonal distortion of the crystal structure leads to significant values of the MAE under particular $c/a$ and alloy concentrations. Here, however, the reduction in MAE is rather small when going from $y=16$ to $y=24$, most likely -- as pointed out above -- because the $c/a$ values are larger for $y=24$ with B than with C, which leads to larger MAEs compared to similar dopant concentrations of C. 
Considering the MAE as a function of Fe concentration $x$ for the case of $y=24$, there is a rather flat behavior, but a maximum value of $0.48~\text{MJ/m}^3$ is obtained for $x=0.35$. For $y=24$ and $x=0.40$ we obtain $\text{MAE}=0.40~\text{MJ/m}^3$, which can be compared to $\text{MAE}=0.19~\text{MJ/m}^3$ obtained with C-doping instead of B. This means that a rather small amount of interstitial B dopants appears to be enough to yield a significant MAE which is promising in a permanent magnet context. 

For systems with $y=24$, we have performed test calculations (not shown) utilizing a simpler approach to the problem of alloying, the virtual crystal approximation (VCA), following the same procedure as described in Ref.~\cite{Delczeg2014}.
Although these calculations overestimate the MAE by a factor between 2 and 4 compared to CPA calculations, nevertheless they result in a similar trend and thus provide a further support to the finding that a substantial MAE can be obtained also for lower Co contents. In contrast to CPA results, VCA ones were obtained within a full potential method.

\section{Experimental results and discussion}

\subsection{Structural properties of Fe-Co-B films}
\begin{figure}
\includegraphics[width=0.5\columnwidth]{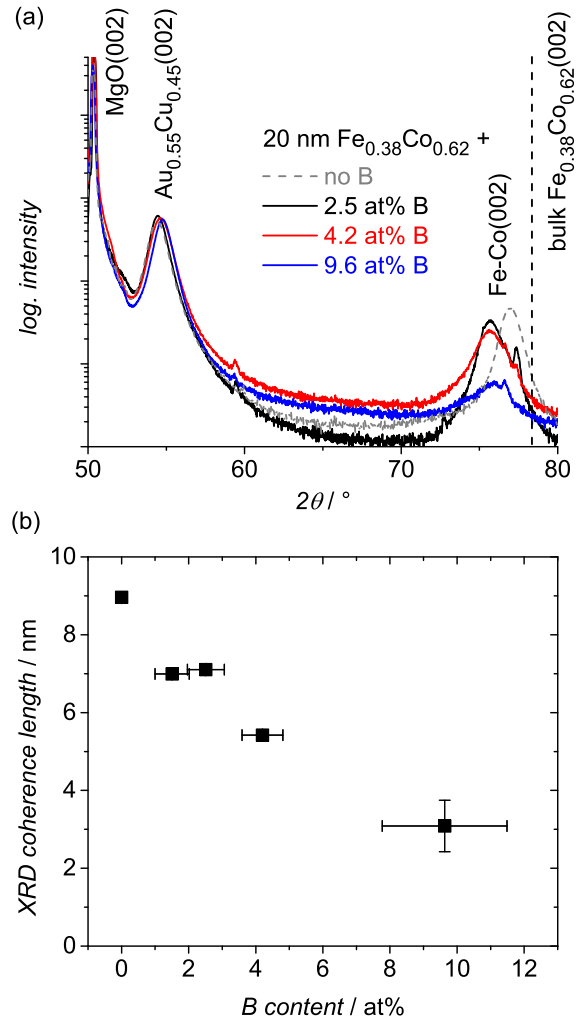}
\caption{\label{XRD}(Color online) (a) XRD patterns of \SI{20}{nm} Fe$_{0.38}$Co$_{0.62}$-B films with varying B content on Au$_{0.55}$Cu$_{0.45}$ buffers. Relevant reflections are marked. The (002) equilibrium 2$\theta$ angle of binary Fe$_{0.38}$Co$_{0.62}$ is given as broken line. The XRD pattern of a Fe$_{0.38}$Co$_{0.62}$ film without B addition is added for comparison. The intensity maxima near 60$^\circ$ and 77$^\circ$ are attributed to PLD droplets with the composition of the used targets. (b) shows the x-ray coherence length according to Scherrer's formula against the AES determined B content.
}
\end{figure}

Using PLD, we prepared a boron composition series and studied the structural and magnetic properties of Fe-Co-B films with \SI{20}{nm} thickness. All films were deposited on Au$_{0.55}$Cu$_{0.45}$ buffer layers, which provide a reduced in-plane lattice parameter compared to Fe-Co and thus induce a $c$ axis oriented film growth~\cite{Kauffmann2014,Reichel2015}. Based on the theoretical results, where the Fe to Co ratio was varied, a composition close to Fe$_{0.4}$Co$_{0.6}$ was chosen as base for our experimental investigations. For the films of this (Fe$_{x}$Co$_{1-x}$)-B series, EDX measurements gave $x=0.38(2)$. Corresponding XRD patterns are shown in Figure~\ref{XRD}(a). The (002) reflection is the only intensity originating from the Fe-Co-B films and indicates an epitaxial film growth of Fe-Co-B on Au-Cu. No additional phases as e.\,g. borides, are detected. Pole figure measurements (see supplementary~\cite{supp}) confirm the epitaxial growth without formation of twins. Compared to the position of the Fe$_{0.38}$Co$_{0.62}$ (002) reflection given in literature~\cite{Predel1993}, we observe a shift to lower 2$\theta$ angles for all samples. This indicates already an expanded $c$ axis of the Fe-Co-B lattice, which will be discussed later in detail. 

The AES depth profiles, which were carried out to determine the B content of the Fe-Co films, did not reveal a significant variation of the B content in dependence on the sputter depth. This allowed us to determine error bars of the B contents for each considered film from its statistical variation. With increasing B content, both a reduced scattering intensity and a peak broadening of the Fe-Co-B(002) reflection is detected. Applying Scherrer's formula, we estimate the x-ray coherence length of the Fe-Co-B films, which can be taken as measure for the Fe-Co-B crystal size. The results are plotted in Figure~\ref{XRD}(b). An increase in B content leads to a substantial decrease of coherence length from almost \SI{10}{nm} in binary Fe$_{0.38}$Co$_{0.62}$ to about \SI{3}{nm} for the film with \SI{9.6}{at\%} B. 

\begin{figure}
\includegraphics[width=0.5\columnwidth]{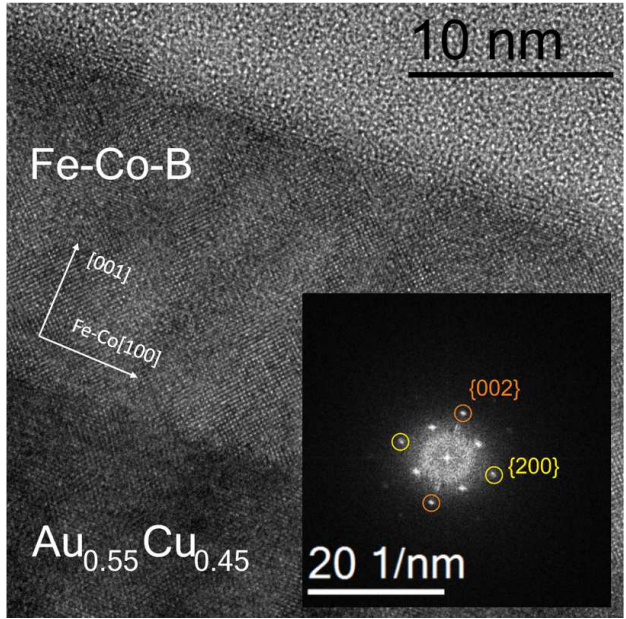}
\caption{\label{TEM}(Color online) TEM image of a \SI{20}{nm} Fe$_{0.38}$Co$_{0.62}$ film with \SI{4.2}{at\%} B deposited on Au$_{0.55}$Cu$_{0.45}$. The zone axis is [010]. The inset shows a FT image of the Fe-Co-B layer. The marked reflections were used for determination of the lattice parameters $c$ and $a$.
}
\end{figure}

A TEM study of a Fe-Co-B film with \SI{4.2}{at\%} B confirms the crystallinity and continuity of the film (Figure~\ref{TEM}). The Fourier Transform (FT) of the Fe-Co-B film (inset) gives a very regular pattern as expected for an epitaxially grown crystal. Minor contrasts within the Fe-Co-B film may indicate the formation of separated grains with a grain size of approx. \SI{5}{nm}. This would be in agreement with the observed x-ray coherence length (Figure~\ref{XRD}(b)), but may also be linked to thickness variations of the lamella.

The strong reduction of crystal size at higher B contents as indicated by the XRD measurements (Fig.~\ref{XRD}(b)) has been confirmed by an additional TEM investigation of the film with the highest B content (see supplementary~\cite{supp}), which revealed oriented nanocrystals of the same diameter surrounded by nanocrystalline and amorphous phases. In literature, both, a formation of Fe-Co-B nanocrystals~\cite{Minor2002} or amorphous Fe-Co-B~\cite{Asai2013} is reported for films of similar B contents. Asai et al. also observed an amorphous structure already at B contents above \SI{5}{at\%} in addition to crystalline Fe-Co-B in Fe$_{0.7}$Co$_{0.3}$-B films~\cite{Asai2013}. Our measurements thus confirm their findings of a decreasing size and fraction of Fe-Co-B crystals with increasing B content in the films. However, all of the presented films contain a significant crystalline fraction which is grown epitaxially. These crystals comprise a certain amount of the alloyed B atoms as is clearly indicated by the change of their lattice parameter $c$, when compared to the binary Fe$_{0.38}$Co$_{0.62}$ film, which is given as reference in Figure~\ref{XRD}(a). 

\begin{figure}
\includegraphics[width=0.5\columnwidth]{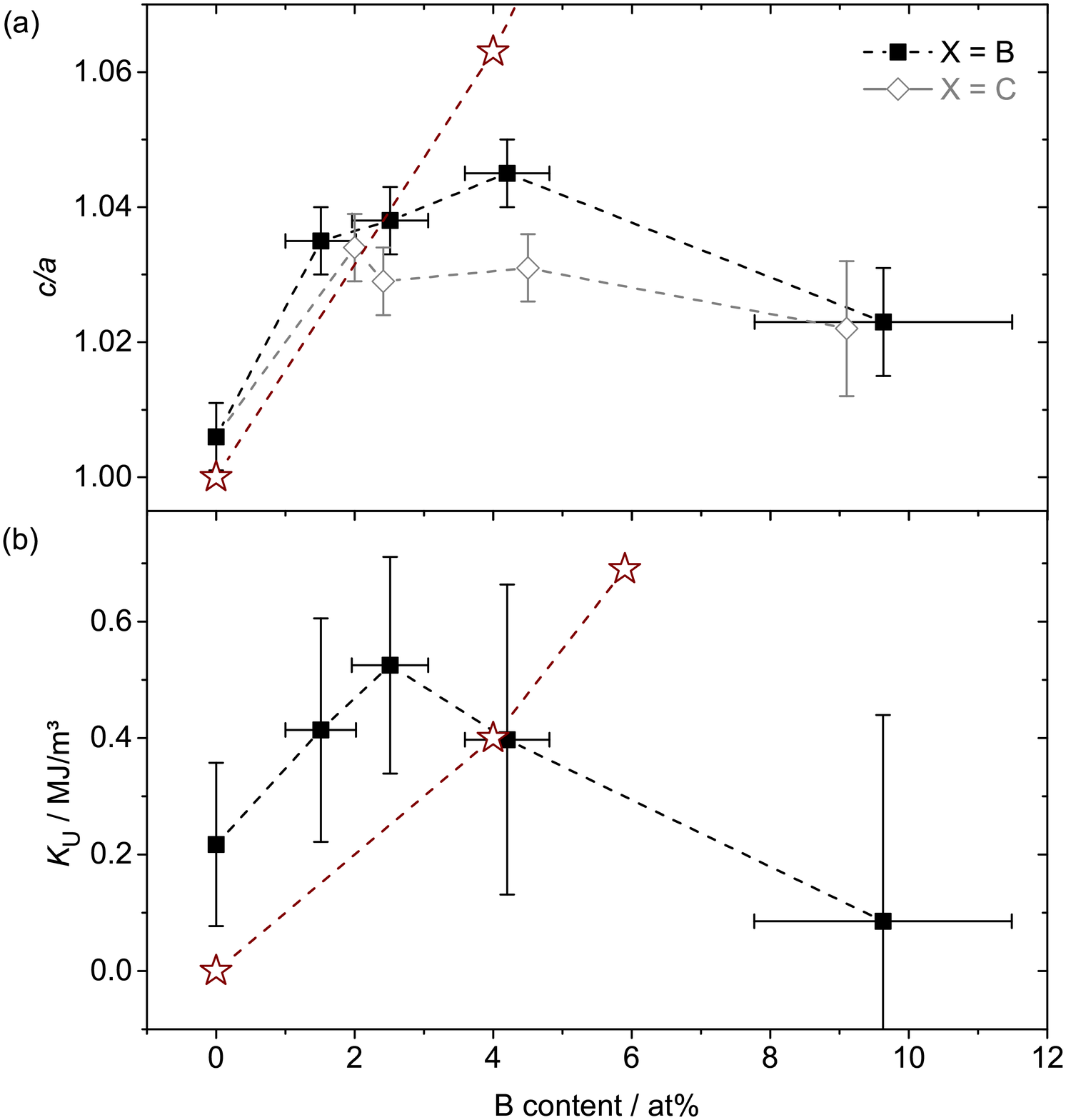}
\caption{\label{distortion}(Color online) (a) Tetragonal distortion $c/a$ for \SI{20}{nm} thick Fe$_{0.38}$Co$_{0.62}$-B films in dependence of the B content (full squares). Results of Fe-Co-C films\cite{Reichel2014} (open diamonds) have been added for comparison. Note that the film with the highest C content is of Fe-C type and did not contain Co. (b) Corresponding uniaxial magnetic anisotropy constants $K_\text{U}$ of the films. The results from the DFT calculations Fe$_{0.4}$Co$_{0.6}$-B (Table~\ref{MAEtable}) have been added as open stars in both graphs. Cubic Fe-Co ($c/a=1$) with no uniaxial MCA ($K_\text{U}=0$) has been added, although it was not treated in our calculations. Broken lines are guides to the eye. 
}
\end{figure}

In order to characterize the tetragonal strain in the epitaxially grown Fe-Co-B films, $\left\lbrace 011\right\rbrace$ pole figures were measured for all films. As described in Ref.\,\cite{Reichel2014}, the $c/a$ ratios were determined from the tilt angles of these planes. In all films, one predominant variant with longer $c$ than in-plane axes was found. No indications for variants with deviating $c$ axis orientation were found. The $c/a$ ratios are plotted in Figure~\ref{distortion}(a) for the different boron contents. An increase of tetragonal distortion is observed with increasing B content up to \SI{4.2}{at\%}. The maximum observed $c/a$ ratio is 1.045 and is achieved for this composition. We can compare this finding to Fe$_{0.4}$Co$_{0.6}$-C films (open diamonds), which exhibit a maximum tetragonal strain of $c/a = 1.03$ and a lower dependence on the C content as discussed in our previous study~\cite{Reichel2014}. The observation of an increased strain in the Fe-Co-B films qualitatively confirms our theoretical calculations, which predicted an enhanced $c/a$ ratio by about 0.03 in Fe-Co-B when compared to Fe-Co-C. The maximum observed strain with $c/a$ of nearly 1.05 is also reasonably close to the predicted $c/a=1.063$ for Fe$_{0.4}$Co$_{0.6}$ with \SI{4}{at\%} B (Table~\ref{MAEtable}). However, a further increase of B content is not followed by a further increased tetragonal strain, which was predicted by DFT. There are different possible reasons for this difference: (a) The limited solubility of boron in Fe-Co which is nearly zero in thermal equilibrium~\cite{Cameron1986}, but is obviously increased substantially by PLD, an effect known for many systems~\cite{Krebs1997}. A limited solubility favors nanocrystalline or amorphous B-rich phases which reduce the effective B content in the Fe-Co matrix and thus also decreases the tetragonal strain. (b) The possible substitutional alloying of B as discussed within the introduction, which would benefit a cubic lattice. And/or (c) the possibility that not all B atoms occupy octahedral interstitials along the axis perpendicular to the film surface, but also interstitials along the in-plane axes are occupied, as e.\,g.\,(1/2;0;0) within the bcc unit cell, which benefit a certain in-plane strain, i.\,e.\,a reduction of the measured $c/a$. 

All reasons have in common that the amount of B atoms, which contribute in a lattice strain along the $c$ axis, is effectively reduced. 
Our DFT calculations were based on the assumption that all B atoms occupy octahedral positions along the $c$ axis. As shown in the calculations in section \ref{preferential}, the B atoms indeed preferentially align in octahedral interstitials along one particular axis, although all three crystal axes are equivalent in a Fe-Co lattice. The experimental results thus confirm these calculations. However, because of the named reasons, not all B atoms contribute to the $c$ axis oriented strain above a B content of \SI{4.2}{at\%}. A likely explanation, why DFT calculations are not able to confirm the experiments exactly is, that they do not consider kinetic effects. Films are prepared within a finite time, which may lead to local variations of B density and thus result in an occupation of different lattice sites than expected from DFT, which describes the ground state. 

The comparison of the B and C doped Fe-Co films~\cite{Reichel2014} implies that a higher amount of B than C atoms can be solved in the PLD prepared Fe-Co films and preferentially occupies interstitial sites along the $c$ axis, which favors a higher strain. An additional contribution may come from the size of the B atoms, which are bigger than C atoms. However, when a certain limit is reached, further added B (or C) atoms lead to a decrease of $c/a$. This limit is reached at around \SI{4}{at\%} B (or C). Since the $c$ axis length is not altered when adding more B or C (see Figure~\ref{XRD}), we conclude that most of the additional atoms occupy sites along the $a$ axes or are dissolved from the Fe-Co crystals. The latter explains the decreasing crystal size with increasing B (or C) content as discussed with Figure~\ref{XRD}(b) and in Ref.\,\cite{Reichel2014}, respectively. 

For the Fe$_{0.38}$Co$_{0.62}$ film with \SI{4.2}{at\%} B, the lattice distortion was determined from the FT of the TEM images (Figure~\ref{TEM}). Compared to the XRD result, where $c/a$ was 1.045 (Figure~\ref{distortion}(a)), we observe a reduced strain of 1.02$\pm$0.01. Such strain reduction after TEM lamella preparation was already reported for Fe-Co-C films and was attributed to a lattice relaxation due to a reduced constraint of a thin TEM lamella compared to a continuous film~\cite{Reichel2014}.

From growth studies of Fe-Co-C films, we concluded that the misfit dislocations which form during film growth affect the whole underlying film~\cite{Reichel2015}. Most of the crystal volume is thus expected to exhibit the same strain state. Only a small fraction of the film close to the buffer interface had an increased strain~\cite{Reichel2014}, expressed by a $c/a$ ratio increased by 0.02. This also holds for the Fe-Co-B films examined here. The tetragonal strain determined close to the Au-Cu buffer from TEM images of the \SI{4.2}{at\%} B containing Fe$_{0.38}$Co$_{0.62}$ film is 1.05$\pm$0.04. This is about 0.03 higher than the value for the film's volume as determined in TEM, which may be explained by the lower in-plane lattice parameter of Au$_{0.55}$Cu$_{0.45}$ compared to Fe-Co~\cite{Reichel2015}. The TEM studies thus prove that only a very small fraction of the films is influenced by this induced strain at the buffer-film interface and most of the film exhibits a spontaneous strain.

The structural characterisation of the films clearly supports our picture of tetragonally strained Fe-Co-B lattices, where the axis perpendicular to the film surface is preferentially strained and the in-plane axes are compressed. The decisive question, why a predominant amount of B atoms preferentially occupies the octahedral interstitials along this axis, which thus is the strained $c$ axis of the tetragonal lattice (Fig.\,\ref{fig_structures}), may be answered as follows. DFT calculations show that there is a strong energetical preference for two nearby B atoms to occupy the same type of octahedral interstitial position. The substrate may act as a \textit{seed layer} to establish that the first B atoms occupy octahedral interstitial in direction perpendicular to the substrate. The DFT results then suggest that such domain will grow preferentially, compared to a random interstitial occupation or formation of multiple domains, both having higher energy. Multiple variants with differently aligned $c$ axes are unfavorable due to the required interface, i.\,e. twin boundary, energy. Thus, Fe-Co-B films exhibit a quasi single crystalline state with one preferentially oriented strained $c$ axis. This axis aligns perpendicularly to the film surface due to the square surface symmetry of the Au-Cu buffer. 

As second main result, the structural measurements revealed spontaneously strained Fe$_{0.38}$Co$_{0.62}$-B phases similar to those in Fe-Co-C~\cite{Reichel2014}. The \SI{20}{nm} thick films are much thicker than coherently strained films. The tetragonal strain is not dependent on film thickness: \SI{100}{nm} thick films with the same compositions (see supplementary~\cite{supp}) exhibit the same $c/a$ ratios as the films presented here. The studied spontaneously strained Fe-Co-B films confirm our DFT calculations, which suggested minima of total energy depending on the B content (Table~\ref{MAEtable}). However, and this is an important difference to the Fe-Co-C films, the spontaneous strain indeed depends on the particular B content -- at least for low B contents up to about \SI{4}{at\%}.

\subsection{Magnetic properties of the Fe$_{0.38}$Co$_{0.62}$-B films}

\begin{figure}
\includegraphics[width=0.5\columnwidth]{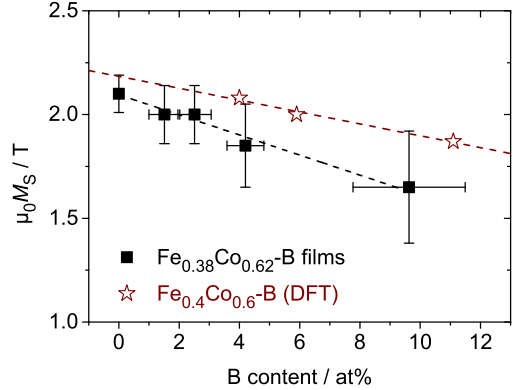}
\caption{\label{Ms}(Color online) Magnetic saturation $\mu_0M_\text{S}$ in dependence of the B content for the Fe$_{0.38}$Co$_{0.62}$-B films (squares) and the corresponding calculated structures (stars) according to Table~\ref{MAEtable}. Linear fits have been added as broken lines.
}
\end{figure}

In order to compare the magnetic properties of spontaneously strained Fe-Co-B films, the anisotropy constant $K_\text{U}$ of the strain related uniaxial MCA was determined from hysteresis measurements as described in Ref.\,\cite{Reichel2014}. The results are summarized in Figure~\ref{distortion}(b). Similar to $c/a$, $K_\text{U}$ also has a maximum in dependence of the B content. The highest MCA with $K_\text{U} = 0.53$\,MJ/m$^3$ is reached in the \SI{2.5}{at\%} B containing Fe$_{0.38}$Co$_{0.62}$ film. Films with smaller B content have a lower MCA because of their reduced tetragonal strain. B contents above \SI{2.5}{at\%} in Fe$_{0.38}$Co$_{0.62}$ do not increase the MCA, although the $c/a$ ratio's maximum is at a B content of \SI{4.2}{at\%} (Figure~\ref{distortion}(a)). The observed decrease of MCA with further increased B content occurs simultaneously with a significant reduction of saturation magnetization in the films (Figure~\ref{Ms}). While our theoretical results described a reduction of $M_\text{S}$ by about \SI{1.5}{\%} with each percent B (open stars), the experiments (squares) reveal a stronger dependence between B content and magnetic saturation: The decrease is about \SI{2.5}{\%} with each \SI{1}{at\%} B. A possible reason might be a decreased Curie temperature due to the addition of B. However, this is not expected~\cite{Hasegawa1978} for B additions in Fe. We rather attribute the higher dependence of $M_\text{S}$ on the B content to a stronger perturbation of the Fe-Co lattice by the B atoms in the PLD prepared films than it is described by the CPA approach, where the B atoms occupy regularly distributed interstitials. The calculations already revealed that Fe and Co atoms have a strongly reduced magnetic moment, when B is their nearest neighbor. In comparison to our experimental study, magnetron sputtered Fe-Co-B films~\cite{Kim2004}, where the material was deposited with substantially lower kinetic energy, exhibit a lower decrease of $M_\text{S}$ with increasing B content, namely \SI{1.5}{\%} per added \SI{1}{at\%} B, which is consistent with our theoretical results (Table~\ref{MAEtable}). 

We argue that both, the strong reduction of $M_\text{S}$ and the decrease of MCA observed in our samples originate from an interplay of the proceeding amorphization due to the increasing B content and the high energy impact of the ions during PLD~\cite{Faehler1996} which may disturb the local environment of the Fe and Co atoms drastically. However, when we compare the theoretical and the experimental results of $K_\text{U}$ (Figure~\ref{distortion}(b)), we find increased values for the Fe$_{0.38}$Co$_{0.62}$-B thin films at low B contents. We attribute this discrepancy to a lack of full potential effects in DFT. At B contents above \SI{4}{at\%}, this underestimation is overcome by the discussed structural effects. We thus observe a very good agreement of the experimentally measured magnetic anisotropy with the theoretically determined value for (Fe$_{0.4}$Co$_{0.6}$)$_{24}$B, but no longer for (Fe$_{0.4}$Co$_{0.6}$)$_{16}$B, i.\,e. \SI{6}{at\%} B (open circles in Figure~\ref{distortion}(b)). 

Concluding the magnetic measurements of the Fe$_{0.38}$Co$_{0.62}$-B films, we observe a direct dependence of the magnetocrystalline anisotropy on the spontaneous strain for low B contents. However, an increased tetragonallity in the Fe-Co-B lattice did not necessarily imply an increased MCA for the complete films due to the formation of non-crystalline or disturbed regions.

\subsection{Structural and magnetic properties in dependence on the Fe/Co ratio}

\begin{figure}
\includegraphics[width=0.5\columnwidth]{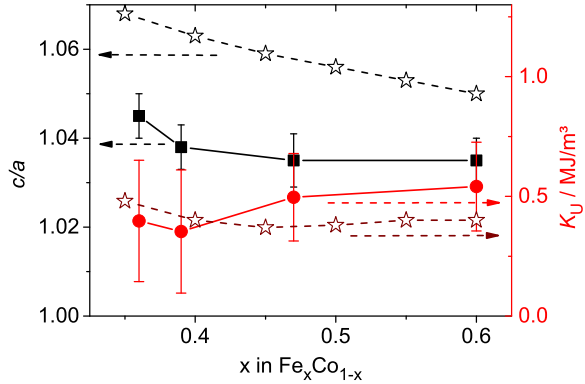}
\caption{\label{series}(Color online) Measured tetragonal lattice strain (full squares) and uniaxial magnetocrystalline anisotropy constants $K_\text{U}$ (full circles) in dependence on Fe-Co ratio in \SI{20}{nm} films with \SI{4}{at\%} B on Au$_{0.55}$Cu$_{0.45}$. DFT results for $c/a$ and $K_\text{U}$ are added (open stars).
}
\end{figure}

The study of various Fe/Co ratios by DFT calculations (Table~\ref{MAEtable}) motivated for a comparison of films with unaltered B content. We chose Fe$_{x}$Co$_{1-x}$ films with \SI{4}{at\%} B due to the observed strain maximum (Figure~\ref{distortion}(a)). The measured tetragonal strain and the observed perpendicular anisotropy constants $K_\text{U}$ in these \SI{20}{nm} thick films are summarized in Figure~\ref{series}. With regard to the tetragonal strain, we observe a slight decrease with increasing Fe content. Although there is only little variation between the films, this finding is qualitatively consistent with the theoretical predictions for (Fe$_x$Co$_{1-x}$)$_{24}$B in Table~\ref{MAEtable}, which also give higher $c/a$ ratios for higher Co contents. Due to already described reasons, we do not reach as high strains as calculated by DFT. The highest experimentally observed $c/a$ is 1.045 for $x = 0.36$. 

When comparing the magnetocrystalline anisotropy energies $K_\text{U}$, we do not observe a strong dependence on the Fe/Co ratio, which in principle fits to the weak dependency proposed by DFT (see Figure~\ref{series}). $K_\text{U}$ is about \SI{0.5}{MJ/m$^3$} on average, but slightly lower on the Co rich side and slightly higher in the films with more Fe than Co. The maximum observed $K_\text{U}$ is \SI{0.54}{MJ/m$^3$} for $x=0.6$. The DFT calculations, however, predicted the highest MCA for the lowest Fe content. 

From the Fe$_{x}$Co$_{1-x}$ films with \SI{4}{at\%} B we thus conclude that higher Fe contents ($x=0.6$) not only lead to a higher magnetic saturation $M_\text{S}$, but are also beneficial for a high MCA at low tetragonal strain. Such Fe rich films or additional concepts to stabilize higher strains beyond $c/a=1.05$ obtained here are considered promising for further research. As one way to fulfill the latter task, we suggest the choice of a preparation method with less energetic impact, where the effect of lattice perturbation due to the B atoms might be smaller. 

\section{Conclusion}

In this combined theory-experimental study, Fe-Co-B is introduced as an alloy with spontaneous strain and hard magnetic properties. Structure calculations based on EMTO predict strong tetragonal lattice distortions up to $c/a = 1.25$ in dependence on the B content, if all B atoms occupy octahedral interstitials along the $c$ axis. This arrangement is preferential, when compared to an occupation of other possible interstitial sites. SPR-KKR results for the distorted Fe-Co-B lattices give strong magnetocrystalline anisotropies when the former cubic symmetry is broken. For \SI{11}{at\%} B, $K_\text{U}$ should be comparable to shape anisotropy. Experiments, however, show that the amount of B atoms which strains the Fe-Co lattice is limited. At a B content higher than \SI{4}{at\%}, the strain decreases. A supersaturation of the octahedral lattice sites with B atoms is considered to be responsible for this deviation from the ideal behavior of B interstitials as treated in DFT. For low B contents, our theoretical results are well confirmed by experiments, in particular when regarding $K_\text{U}$. The highest magnetocrystalline anisotropies are observed for Fe$_{0.38}$Co$_{0.62}$ with \SI{2.5}{at\%} B or in Fe richer Fe$_{0.6}$Co$_{0.4}$ with \SI{4}{at\%} B. $K_\text{U}$ of these alloys is above \SI{0.5}{MJ/m$^3$}. Although this value does not exceed the shape anisotropy, the spontaneous strain should allow for a preparation of Fe-Co-B with an easy axis of magnetization, which is defined by structure and not by shape. The approach to exploit spontaneous strain and the related MCA thus gives much more opportunities than the substrate induced strain in Fe-Co, where the strain is limited to ultrathin films.

\acknowledgements
We acknowledge funding of the EU through FP7-REFREEPERMAG. For TEM lamella preparation, we thank Christine Damm, Juliane Scheiter and Almut Pöhl. The authors further thank Levente Vitos, Christian Behler and Ruslan Salikhov for discussion. Calculations were performed on UPPMAX,  NSC-Matter, Triolith and C3SE resources. E.\,K.\,Delczeg-Czirjak and J.\,Rusz acknowledge the Swedish Research Council for financial support.


\begin{thebibliography}{10}
\bibitem{Bcontent}In this study, experimentally determined B contents are given in at\%, although the DFT calculations deal with (ideal) supercell structures of (Fe-Co)$_y$B, where B occupies octahedral interstitial sites.
\end{thebibliography}
\end{document}